%% file: main.tex
\documentclass[nonacm, screen]{acmart}
\usepackage{amsmath}
\DeclareMathOperator*{\argmin}{arg\,min}
\newtheorem{definition}{Definition}
\newtheorem{exmp}{Example}
\usepackage{macros}
\usepackage{subfig}
\usepackage[title]{appendix}
\AtBeginDocument{%
  \providecommand\BibTeX{{%
    \normalfont B\kern-0.5em{\scshape i\kern-0.25em b}\kern-0.8em\TeX}}}
    
\usepackage{titlesec}
\titlespacing*{\subsection}
{0pt}{1ex}{0ex}
\titlespacing*{\section}
{0pt}{1ex}{0ex}
\titlespacing*{\paragraph}
{0pt}{1ex}{0ex}
\acmSubmissionID{123-A56-BU3}

\newcommand{\secref}[1]{Section \ref{#1}}

\newcommand{\figref}[1]{Fig. \ref{#1}}

\newcommand{\tabref}[1]{Table \ref{#1}}

\newcommand{\commentout}[1]{%
}

\newcommand\PSL{%
  \textsc{PSL}%
}

\newcommand\HyperFair{%
  \textsc{HyperFair}%
}

\newcommand\MovieLens{%
  \textsc{MovieLens}%
}

\newcommand\Hyper{%
  \textsc{HyPER}%
}

\setcopyright{none}
\authorsaddresses{}
\begin{document}

\title{\HyperFair: A Soft Approach to Integrating Fairness Criteria}


\author{Charles Dickens}
\affiliation{%
  \institution{University of California, Santa Cruz}
  \streetaddress{1156 High Street}
  \city{Santa Cruz}
  \country{United States of America}
  }

\author{Rishika Singh}
\affiliation{%
  \institution{University of California, Santa Cruz}
  \streetaddress{1156 High Street}
  \city{Santa Cruz}
  \country{United States of America}
  }

\author{Lise Getoor}
\affiliation{%
  \institution{University of California, Santa Cruz}
  \streetaddress{1156 High Street}
  \city{Santa Cruz}
  \country{United States of America}
  }

\begin{abstract}
    \input{sections/abstract}
\end{abstract}

\maketitle

\input{sections/introduction}


\input{sections/background}
\input{sections/models}

\section{Empirical Evaluation}
\label{sec:experiments}
\input{sections/experiments}

\section{Conclusion and Future Work}
\label{sec:conclusion}
\input{sections/conclusion}

\section{Acknowledgements}
\label{sec:acknowledgements}
\input{sections/acknowledgements}

\bibliography{fair.bib}
\bibliographystyle{plain}

\begin{appendices}
\input{sections/appendix}

\end{appendices}

\end{document}

%% file: sections/abstract.tex
\footnotesize{
Recommender systems are being employed across an increasingly diverse set of domains that can potentially make a significant social and individual impact.
For this reason, considering fairness is a critical step in the design and evaluation 
of such systems.
In this paper, we introduce \HyperFair, a general framework for enforcing soft fairness constraints in a hybrid recommender system.
\HyperFair \, models integrate variations of fairness metrics as a regularization of a joint inference objective function.
We implement our approach using probabilistic soft logic and show that it is particularly well-suited for this task as it is expressive and structural constraints can be added to the system in a concise and interpretable manner.
We propose two ways to employ the methods we introduce: first as an extension of a probabilistic soft logic recommender system template;
second as a fair retrofitting technique that can be used to improve the fairness of predictions from a black-box model.
We empirically validate our approach by implementing multiple \HyperFair \, hybrid recommenders and compare them to a state-of-the-art fair recommender. %
We also run experiments showing the effectiveness of our methods for the task of retrofitting a black-box model and the trade-off between the amount of fairness enforced and the prediction performance.
}

%% file: sections/introduction.tex
\section{Introduction}
\label{sec:intro}


As the ubiquity of recommender 
systems
continues to grow, concerns of bias and fairness are becoming increasingly urgent to address.
An algorithm oblivious to any form of fairness
has the potential to propagate, or even amplify, discrimination \cite{MenAlsoLikeShopping:Zhao, DiscriminationAds}. 
In doing so, certain groups can be severely impacted by the recommendations provided.
For instance, one study showed that an algorithm for targeted advertising of jobs in the STEM fields was delivering more advertisements to men than women with similar professional backgrounds \cite{Lambrecht:biasedAdvertising}.
The need to integrate fairness and ensure that different groups of users are experiencing the same level of utility from recommender systems has been acknowledged 
by the artificial intelligence community \cite{mehrabi2019survey, pmlr-v81-ekstrand18b}. 
We introduce techniques for integrating fairness metrics
as regularizations of a joint inference objective function of a probabilistic graphical model.
Our approach naturally leads to novel collections of rules that can be added to a probabilistic soft logic (\PSL) \cite{bach:jmlr17} model. 
Furthermore, the weights of the rules can be translated as regularization parameters which can be tuned by the modeler or via weight learning \cite{bach:jmlr17}.
This motivates a general framework for introducing multiple soft fairness constraints in a hybrid recommender system which we refer to as \HyperFair.
\HyperFair \, builds upon the \Hyper \, recommender system introduced by Kouki et al. \cite{kouki:recsys15} by adding the ability to enforce multiple soft fairness constraints to the model predictions. 
This framework is general enough to capture previous work by Farnadi et al. \cite{fairHyper:RecSys} who proposed \PSL \, modelling techniques for addressing disparities stemming from imbalanced training data and observation bias.
We develop a generic technique, provide principled derivations of the soft constraints, and show how a set of fairness metrics can be precisely targeted.   


Our key contributions are as follows: 1) we introduce the \HyperFair \, framework for enforcing soft fairness constraints in a hybrid recommender system; 2) we show that non-parity and value unfairness can be written as linear combinations of hinge-loss potentials and can thus be integrated into the \PSL \, inference objective via template rules;
3) we perform an empirical analysis using the \MovieLens\, dataset and show both how our fairness rules can be used
internally or as a method for retrofitting the output of a black-box algorithm to increase the fairness of predictions; 
and 4) we show our method improves fairness over baseline models and outperforms a state-of-the-art fair recommender \cite{DBLP:journals/corr/YaoH17} in terms of RMSE and value unfairness.

\newpage

%% file: sections/background.tex
\section{Background}
\label{sec:background}

We begin by briefly reviewing related work upon which our approach builds.


\subsection{Fairness in Recommender Systems}

Methods for addressing fairness can occur at three stages of a recommender pipeline: pre-process, in-process, and post-process \cite{mehrabi2019survey}.
Pre-processing techniques transform the data so that discrimination characteristics are removed prior to model training \cite{lahoti2019ifair, NIPS2017_6988}.
In-processing techniques attempt to remove discrimination during the model training process by incorporating changes into the objective function.
Post-processing techniques treat the learned model as a black box and modify the output to remove discrimination \cite{FairnessExposureInRankings:Singh,  equalityOfOpportunity:Hardt, LinkedInReRanking:Geyik, MatchmakingFairness:Paraschakis, fair-Reranking:recsys:2019}.
Post-processing techniques are particularly attractive to industry since their treatment of the predictor as a black-box makes for a manageable integration into an existing pipeline
\cite{AdressingAlgorithmicBias:Gathright, LinkedInReRanking:Geyik}.
Recent work has shown the effectiveness of both adversarial learning
\cite{AdversariallyLearningFairRepresentations:Beutel, Louizos:CoRR2016, Madras:CoRR2018} 
and regularization \cite{FairnessAwareRegularizer, DBLP:journals/corr/YaoH17, PairwiseFairness}. 
Our methods can be used as either an in-processing or post-processing method, and
builds upon the line of research that addresses fairness via regularization.

\subsection{Hybrid Recommenders using Probabilistic Soft Logic}
\label{sec:Hyper}

Probabilistic soft logic (\PSL) is a probabilistic programming language that 
has been shown to be effective for hybrid recommender systems \cite{kouki:recsys15}.  \PSL's advantages include the ability to easily write interpretable, extendable, and explainable hybrid systems \cite{kouki:recsys17}. 
\PSL \, models specify probabilistic dependencies using logical and arithmetic rules; the rules, combined with data, are translated into a conditional random field referred to as a \emph{hinge-loss Markov random field (HL-MRF)} \cite{bach:jmlr17}.   
Given a set of evidence $\mathbf{x}$ and continuous unobserved variables $\mathbf{y}$, the inference objective is given by:
\vspace{-1mm}
\begin{equation}
    \min_{\mathbf{y} \in [0, 1]^n} \quad
        \sum_{i}^{k} w_{i} \phi_{i}(\mathbf{y}, \mathbf{x})
    \label{eq:Rec_Obj}
\end{equation}
\noindent
where $k$ is the number of unique hinge-loss potential functions, $\phi_i(\cdot)$, and $w_i$ are the corresponding scalar weights.
In addition to expressivity, an important advantage of HL-MRFs is their scalability; inference is convex, and a variety of specialized optimizers have been proposed \cite{srinivasan:aaai20} (more details about \PSL \, in \secref{sec:appendix}). 

Following Kouki et al. \cite{kouki:recsys15}, the following collection of rules expresses a simple, intuitive hybrid recommender model:

\paragraph{\bf Demographic and Content Similarity:}
\, Demographic-based approaches are built upon on the observation that users with similar demographic properties will tend to make similar ratings. 
Likewise, content-based approaches are built upon on the observation that items with similar content will be rated similarly by users.
This is different from the collaborative filtering approach as the rating patterns of users and items is strictly not considered in the similarity calculation.
\vspace{-1mm}
\begin{gather*}
\pslpred{Rating}(\pslarg{U1}, \pslarg{I})
\psland
\pslpred{SimUserDemo}(\pslarg{U1}, \pslarg{U2})
\pslthen
\pslpred{Rating}(\pslarg{U2}, \pslarg{I})
\\
\pslpred{Rating}(\pslarg{U}, \pslarg{I1})
\psland
\pslpred{SimItemContent}(\pslarg{I1}, \pslarg{I2})
\pslthen
\pslpred{Rating}(\pslarg{U}, \pslarg{I2})
\end{gather*}
\vspace{-1mm}
The predicate $\pslpred{Rating}(\pslarg{U}, \pslarg{I})$ represents the normalized value of the rating that user $\pslarg{U}$ provided for item $\pslarg{I}$.\\
$\pslpred{SimUserDem}(\pslarg{U1}, \pslarg{U2})$ and $\pslpred{SimItemContent}(\pslarg{I1}, \pslarg{I2})$ represent the similarity of users $\pslarg{U1}$ and $\pslarg{U2}$ and items $\pslarg{I1}$ and $\pslarg{I2}$.

\paragraph{\bf Neighborhood-based Collaborative Filtering:}
\, Neighborhood-based collaborative filtering methods capture the notion that users that have rated items similarly in the past will continue to rate new items similarly.  
An analogous and transposed notion applies to items, i.e., items that have been rated similarly by many of the same users will continue to be rated similarly.
Similarity in this context is based solely on rating patterns and can be measured using various metrics.
\vspace{-1mm}
\begin{gather*}
\pslpred{Rating}(\pslarg{U1}, \pslarg{I})
\psland
\pslpred{SimUsers}(\pslarg{U1}, \pslarg{U2})
\pslthen
\pslpred{Rating}(\pslarg{U2}, \pslarg{I})
\\
\pslpred{Rating}(\pslarg{U}, \pslarg{I1})
\psland
\pslpred{SimItems}(\pslarg{I1}, \pslarg{I2})
\pslthen
\pslpred{Rating}(\pslarg{U}, \pslarg{I2})
\end{gather*}
\vspace{-1mm}
$\pslpred{SimUsers}(\pslarg{U1}, \pslarg{U2})$ and $\pslpred{SimItems}(\pslarg{I1}, \pslarg{I2})$ represent the similarity of users $\pslarg{U1}$ and $\pslarg{U2}$ and items $\pslarg{I1}$ and $\pslarg{I2}$, respectively. 

\paragraph{\bf Local Predictor Prior:} \,
One of the advantages of the \Hyper \, system is its ability to combine multiple recommendation algorithms into a single model in a principled fashion.
Recommender predictions are incorporated as non-uniform priors in the \PSL \, model using the pattern shown below.
\begin{gather*}
\pslpred{LocalPredictor}(\pslarg{U}, \pslarg{I})
=
\pslpred{Rating}(\pslarg{U}, \pslarg{I})
\end{gather*}
The predicate $\pslpred{LocalPredictor}(\pslarg{U}, \pslarg{I})$ represents the prediction made by the external recommendation algorithm. 

\paragraph{\bf Mean-Centering Priors:}
%
\, Based on the above rules, ratings are propagated across similar users, and if two users have different average ratings, then these ratings may actually bias one another too much.
To counter this effect, the following rules bias ratings towards the average user and item rating calculated from the observed ratings:
\begin{gather*}
\pslpred{AverageUserRating}(\pslarg{U})
=
\pslpred{Rating}(\pslarg{U}, \pslarg{I})
\\
\pslpred{AverageItemRating}(\pslarg{I})
=
\pslpred{Rating}(\pslarg{U}, \pslarg{I})
\end{gather*}
The predicates $\pslpred{AverageItemRating}(\pslarg{I})$ and $\pslpred{AverageUserRating}(\pslarg{U})$ represent the average normalized value of the ratings associated with user $\pslarg{U}$ and item $\pslarg{I}$, respectively. 


\subsection{FairPSL}

Farnadi et al. \cite{fairHyper:RecSys} introduced two collections of rules that can be added to a \PSL \, recommender system to address disparities derived from training data and observations.
The authors use the same metrics we consider as proxies to measure this notion of disparity.
Our work is different in that we give the modeler a finer level of control.
Rather than attempting to capture multiple notions of fairness at once, we propose techniques that integrate specific fairness metrics into the \PSL\, inference objective as regularizers.
In this way, the modeler is able to tune the degree of the specific metric to the domain they are working in simply by adjusting the weight of the additional rules.  

%% file: sections/models.tex
\section{\HyperFair}
\label{sec:models}

In this section, we introduce \HyperFair, a framework for enforcing multiple soft fairness constraints in a hybrid recommender system.
\HyperFair \, is a natural development to \Hyper \cite{kouki:recsys15} that incorporates fairness metrics, $U$, via regularization of the HL-MRF MAP inference objective \eqref{eq:Rec_Obj}:
\begin{equation}
    \min_{\mathbf{y} \in [0, 1]^n} \quad
        w_f U + \sum_{i}^{k} w_{i} \phi_{i}(\mathbf{y}, \mathbf{x})
    \label{eq:regularized_rec_obj}
\end{equation}
where $w_f \in \mathcal{R}^+$ is the scalar regularization parameter.
A fairness metric, $U$, in a \HyperFair \, model is expressed as a linear combination of hinge loss potential functions and can then be written as a \PSL \, rule.

A particularly active and productive area of research is defining fairness metrics, and there are many metrics that could be targeted by the proposed HyperFair framework.
Following the line of work in \cite{DBLP:journals/corr/YaoH17} and \cite{fairHyper:RecSys}, we focus on the unfairness metrics of non-parity and value unfairness defined in the following sections.
These metrics were introduced in \cite{DBLP:journals/corr/YaoH17} specifically for addressing disparity stemming from biased training data in collaborative-filtering based recommender systems.

For the remainder of this paper, we will let $g$ represent the protected group of users, i.e., $g$ is a subset of all the users present in the data that have been identified as possessing a protected attribute.
Then, $\neg g$ represents the remaining subset of users that do not possess the protected attribute.
We let $\textbf{R}$ be the set of ratings in the dataset, $m$ the number of predictions made by the model, $n$ the number of unique items in the dataset, and $v_{i,j}$ and $r_{i,j}$ the predicted and true rating user $i$ gives item $j$, respectively. 
For the fairness metric definitions, we use $E_{g}[v]$ and $E_{\neg g}[v]$ to represent the average predicted ratings for $g$ and $\neg g$, respectively, $E_{g}[v]_{j}$ and $E_{\neg g}[v]_{j}$ represent the average predicted ratings for item $j$ for $g$ and $\neg g$, respectively, and $E_{g}[r]_{j}$ and $E_{\neg g}[r]_{j}$ represent the average true ratings for item $j$ for $g$ and $\neg g$, respectively.


\subsection{Non-parity Unfairness}



Non-parity unfairness, $U_{par}$, 
aims to minimize the disparity in the overall average predicted ratings of the protected and unprotected groups. 

$$U_{par}(v) = |(E_{g}[v] -  E_{\neg g}[v])|$$

In this section, we motivate a collection of rules that can be added to the \PSL \, recommender system as an approach to minimize this metric.
We start by introducing two new free variables to the inference problem \eqref{eq:Rec_Obj}, $y_{n + 1}$ and $y_{n + 2}$, and the following hard constraints without 
breaking the convexity of \PSL \, inference:
\begin{gather*}
    c_{1}(\mathbf{y}, \mathbf{x}) := 
    y_{n + 1} - \frac{1}{\lvert \{(i,j) : ((i, j) \in \mathbf{R}) \land g_{i} \}\rvert} 
    \sum_{(i,j) : ((i, j ) \in \mathbf{R}) \land g_{i}} v_{i,j}
    = 0 \\
    c_{2}(\mathbf{y}, \mathbf{x}) := 
    y_{n + 2} - \frac{1}{\lvert \{(i,j) : ((i, j ) \in \mathbf{R}) \land \neg g_{i} \}\rvert} 
    \sum_{(i,j) : ((i, j ) \in \mathbf{R}) \land \neg g_{i}} v_{i,j}
    = 0
\end{gather*}
With the two additional hard constraints, the solution of the new optimization problem is 
a state such that $y^{*}_{n + 1} = E_{g}[v]$ and $y^{*}_{n + 2} = E_{\neg g}[v]$. 
The two hard constraints can be added to the \PSL \, model by introducing the following pair of rules:
\begin{gather*}
\pslpred{Rating}(+ \pslarg{U}, + \pslarg{I}) / m_{g}
=
\pslpred{ProtectedAvgRating}(\pslarg{c}) \, . \, 
\{\pslarg{U}: \pslpred{Protected}(\pslarg{U})\} \, \{\pslarg{I}: \pslpred{ProtectedItem}(\pslarg{I})\} \\ 
\pslpred{Rating}(+ \pslarg{U}, + \pslarg{I}) / m_{\neg g}
=
\pslpred{UnProtectedAvgRating}(\pslarg{c}) \, . \,
\{\pslarg{U}: \pslpred{UnProtected}(\pslarg{U})\} \, \{\pslarg{I}: \pslpred{UnProtectedItem}(\pslarg{I})\} 
\end{gather*}
where $m_{g}$ and $m_{\neg g}$ are the total number of ratings for the protected and unprotected group, respectively, and are added as a preprocessing step. 
The predicates $\pslpred{ProtectedAvgRating}(\pslarg{c})$ and $\pslpred{UnProtectedAvgRating}(\pslarg{c})$ hold the values of $y_{n+1}$ and $y_{n+2}$
, respectively.
We can now define the two following hinge-loss potentials:
\begin{align*}
\phi_{k+1} (\mathbf{y}, \mathbf{x}) = \max \Big \{1 - y_{n+1} - (1 - y_{n + 2}), 0 \Big\} &&
\phi_{k+2}(\mathbf{y}, \mathbf{x}) = \max \Big \{1 - y_{n+2} - (1 - y_{n + 1}), 0 \Big\}
\end{align*}
\noindent
Observe that $U_{par} = \phi_{k+1} (y_{n+1}^{*}, \mathbf{x}) + \phi_{k+2} (y_{n+2}^{*}, \mathbf{x})$.
This transformation allows us to push the regularizer in \eqref{eq:regularized_rec_obj} into the summation to create a valid \PSL \, objective function.
Formally, if we let $w_{k + 1} = w_{k + 2} = w_f$, then:
\begin{align}
    \argmin_{\mathbf{y} \in [0, 1]^n} \quad
    w_f U_{par} + \sum_{i}^{k} w_{i} \phi_{i}(\mathbf{y}, \mathbf{x}) \quad \equiv \quad \argmin_{\mathbf{y} \in [0, 1]^{n + 2}} \quad &  
    \sum_{i}^{k + 2} w_{i} \phi_{i}(\mathbf{y'}, \mathbf{x}) \label{eq:NP_Rec_Intervention_Obj} \\
     \textrm{s.t.} \quad & c_{1}(\mathbf{y}, \mathbf{x}) = 0, \  c_{2}(\mathbf{y}, \mathbf{x}) = 0 \nonumber 
\end{align}
\noindent
Furthermore, we now have that the right hand side of \eqref{eq:NP_Rec_Intervention_Obj} is a valid HL-MRF that can be instantiated using \PSL.

The following rule can be added to a \PSL \, model to obtain the two desired ground potentials $\phi_{k+1}$ and $\phi_{k+2}$.
\begin{gather*}
\pslpred{ProtectedAvgRating}(\pslarg{c})
=
\pslpred{UnProtectedAvgRating}(\pslarg{c})
\end{gather*}
Altogether, this method for addressing non-parity unfairness results in a total of $4$ additional ground potentials and $3$ additional rules in the \PSL \, template and achieves precisely the desired semantics.
Furthermore, the regularization term $w_f$ is directly translated as a weight in \PSL \, that can be tuned by the modeler or via weight learning.

\subsection{Value Unfairness}
Next, we motivate our approach in addressing value unfairness. 
Value unfairness aims to minimize the expected inconsistency in the signed estimation error between the protected and unprotected user groups.
$$U_{val}(y) = \frac{1}{n} 
\sum_{j=1}^{n} |(E_{g}[v]_{j} - E_{g}[r]_{j}) 
- (E_{\neg g}[v]_{j} - E_{\neg g}[r]_{j})|$$
A key difference between this metric and non-parity is that the truth values of the predictions are included in the definition of the metric and thus cannot be directly targeted during inference.
In \PSL \, the truth values of the target predicates are withheld until evaluation. 
Therefore, the approach we take is to estimate properties of the rating distribution prior to running the model to approximate the desired inference objective function \eqref{eq:regularized_rec_obj}.


We start the derivation of the fairness rules we will be adding to the \PSL \, model by augmenting the inference optimization problem of \eqref{eq:Rec_Obj} with two hard constraints for every item in the dataset, that is for all $j \in I = \{j : (i,j) \in \mathbf{R}\}$.
Note that these constraints do not break the convexity properties 
of the original optimization problem.
\begin{gather*}
    c_{1,j}(y_{n + j}, \mathbf{x}) := 
    y_{n + j} - \frac{1}{\lvert \{i : ((i, j) \in \mathbf{R}) \land g_{i} \}\rvert}
    \sum_{i : ((i, j ) \in \mathbf{R}) \land g_{i}} v_{i,j}
    = 0 \\
    c_{2,j}(y_{n + j + \lvert I \rvert}, \mathbf{x}) := 
    y_{n + j + \lvert I \rvert} - \frac{1}{\lvert \{i : ((i, j) \in \mathbf{R}) \land \neg g_{i} \}\rvert}
    \sum_{i : ((i, j) \in \mathbf{R}) \land \neg g_{i}} v_{i,j}
    = 0
\end{gather*}
With these hard constraints, the setting of the free variables $y_{n + 1}, \cdots , y_{n + \lvert I \rvert}, y_{n + \lvert I \rvert} \cdots y_{n + 2 \lvert I \rvert}$ in the optimal solution will be such that 
$(y^{*}_{n + 1} = E_g[v]_1), \cdots , (y^{*}_{n + \lvert I \rvert} = E_g[v]_{\lvert I \rvert}), (y^{*}_{n + 1 + \lvert I \rvert} = E_{\neg g}[v]_{1}) \cdots (y^{*}_{n + 2 \lvert I \rvert} = E_{\neg g}[v]_{ \lvert I \rvert})$. 
These hard constraints are added to the \PSL \, inference objective with the following rule:
\begin{gather*}
\pslpred{Rating}(+ \pslarg{U}, \pslarg{I}) / @Max[1, \lvert \pslarg{U} \rvert]
=
\pslpred{PredGroupAvgItemRating}(\pslarg{G}, \pslarg{I}) \, . \, 
\{\pslarg{U}: \pslpred{target}(\pslarg{U}, \pslarg{I}) \land \pslpred{Group}(\pslarg{G}, \pslarg{U})\}
\end{gather*}
$\pslpred{PredGroupAvgItemRating}(\pslarg{G}, \pslarg{I})$ represents the average of the predicted ratings that users in group $G$ gave item $I$.
The term $G$ in this rule represents either the unprotected or protected group.


At the time of inference, we cannot calculate the average true values of the ratings for either the protected or unprotected group, $E_{g}[r]_{j}$ or $E_{\neg g}[r]_{j}$, since the true rating value information is withheld until evaluation. 
Instead, the group average item rating is estimated using the observed ratings, $\hat{E}_{g}[r]_{j} = \frac{1}{\lvert \{i : ((i, j) \in \mathbf{R}_{obs}) \land g_{i} \}\rvert} \sum_{i : ((i, j) \in \mathbf{R}_{obs}) \land g_{i}} v_{i,j}$ and similarly, $\hat{E}_{\neg g}[r]_{j} = \frac{1}{\lvert \{i : ((i, j) \in \mathbf{R}_{obs}) \land \neg g_{i} \}\rvert} \sum_{i : ((i, j) \in \mathbf{R}_{obs}) \land \neg g_{i}} v_{i,j}$, where $\mathbf{R}_{obs}$ is the set of observed ratings. 
The observed group average item rating is calculated in a preprocessing step and is added to the model as an observed predicate, $\pslpred{ObsGroupAvgItemRating}(\pslarg{G}, \pslarg{I})$. 

We can now define the following set of hinge-loss potentials:
\begin{align*}
\phi_{k + j}(\mathbf{y}, \mathbf{x}) & = \max\Big \{(y_{n + j} - \hat{E}_{g}[r]_{j}) - (y_{n + j + \lvert I \rvert} - \hat{E}_{\neg g}[r]_{j}), 0 \Big \} \\
\phi_{k + j + \lvert I \rvert}(\mathbf{y}, \mathbf{x}) & = \max\Big \{(y_{n + j + \lvert I \rvert} - \hat{E}_{\neg g}[r]_{j}) - (y_{n + j} + \hat{E}_{g}[r]_{j}), 0 \Big \}
\end{align*}
Then, in the optimal state: $U_{val} \approx n \sum_{j = 1}^{2 \lvert I \rvert} \phi_{k + j} (\mathbf{y}^{*}, \mathbf{x}) =: \hat{U}_{val}$.
This transformation allows us to push an approximation of the regularizer in \eqref{eq:regularized_rec_obj} into the summation of the HL-MRF MAP inference objective function.
Formally, if we let $w_{k+j} = w' = \frac{1}{n} w_{Val}$ for all $j \geq 0$, then:
\begin{align}
    \argmin_{\mathbf{y} \in [0, 1]^n} \quad
    w_f \hat{U}_{val} + \sum_{i}^{k} w_{i} \phi_{i}(\mathbf{y}, \mathbf{x}) \quad \equiv
     \argmin_{\mathbf{y'} \in [0, 1]^{n + 2|I|}} \quad &
        \sum_{i}^{k + 2|I|} w_{i} \phi_{i}(\mathbf{y'}, \mathbf{x}) \label{eq:val_Rec_Intervention}\\
         \textrm{s.t.} \quad & c_{1, j}(\mathbf{y'}, \mathbf{x}) = 0, \ 
         c_{2, j}(\mathbf{y'}, \mathbf{x}) = 0, \ \forall j \in I \nonumber
\end{align}
Further, we have \eqref{eq:val_Rec_Intervention} is a valid HL-MRF that can be instantiated using \PSL.
Specifically, the following rule in \PSL \, results in the desired potentials $\phi_{k + 1}, \cdots, \phi_{k + 2 \lvert I \rvert}$:
\begin{align*}
    \pslpred{PredGroupAvgItemRating} (\pslarg{G1}, \pslarg{I}) & - \pslpred{ObsGroupAvgItemRating} (\pslarg{G1}, \pslarg{I}) \\ = 
    \pslpred{PredGroupAvgItemRating} (\pslarg{G2}, \pslarg{I}) & - \pslpred{ObsGroupAvgItemRating} (\pslarg{G2}, \pslarg{I})
\end{align*}
This approach approximates the targeted fairness metric using statistics from the set of observations.
The approximation is then transformed into a summation of hinge-loss potential functions that could be pushed into a \PSL \, inference objective function.
Furthermore, the weight of the arithmetic rule in this intervention can be interpreted as a scaled version of the regularization parameter of the fairness metric in \eqref{eq:regularized_rec_obj}.

%% file: sections/experiments.tex


We evaluate our proposed \PSL \, fairness interventions on the Movielens 1M dataset \cite{harper2015movielens}.
In addition to the ratings, this dataset includes auxiliary information such as movie metadata (e.g., genres, title, and release date) and user demographic information (e.g., gender, age, and occupation).
In table 2 of \cite{DBLP:journals/corr/YaoH17}, the authors summarize some of the gender-based statistics of the \MovieLens \, 1M dataset which underscores the disparity in the ratings. 
Therefore, following the same preprocessing steps taken in previous related work \cite{DBLP:journals/corr/YaoH17,fairHyper:RecSys}, the protected and unprotected groups are chosen to be female and male, respectively, and movies are filtered by genre, considering only those with \textit{action}, \textit{romance}, \textit{crime}, \textit{musical}, and \textit{sci-fi} tags, and finally users with fewer than $50$ ratings are removed.
The remaining dataset contains approximately 450K timestamped ratings made by 3K users across 1K movies.

\subsection{In-Process Fairness Intervention}

The baseline hybrid recommender system (\secref{sec:Hyper})
uses cosine similarity for the similarity predicates and
three local predictors, non-negative matrix factorization (NMF) \cite{lee2001algorithms}, biased singular value decomposition (SVD) \cite{koren2009matrix}, and a content-based multinomial Naive Bayes multi-class classifier with Laplace smoothing that is trained using the demographic and content information of the user and item, respectively.
Five versions of the \PSL \, recommender system, extending the baseline model, are implemented:

\commentout{
The baseline hybrid recommender system discussed in \secref{sec:Hyper} is built with the following implementation details.
The similarity predicates all represent calculated cosine similarities.
Furthermore, three local predictors are employed.
The first two are the matrix factorization based approaches, non-negative matrix factorization (NMF) \cite{lee2001algorithms} and biased singular value decomposition (SVD) \cite{koren2009matrix}.
The third local predictor is a content-based multinomial Naive Bayes multi-class classifier with Laplace smoothing that is trained using the demographic and content information of the user and item, respectively.
Five versions of the \PSL \, recommender system, extending the baseline model, are implemented:
}
\begin{itemize}
    \item \textbf{\PSL \, Base:} demographic and content similarity, collaborative filtering, local predictor, and mean-centering rules
    
    \item \textbf{\HyperFair (NP):} The \textbf{\PSL \, Base} model with the non-parity fairness rules.
    
    \item \textbf{\HyperFair (Val):} The \textbf{\PSL \, Base} model with the value fairness rules.

    \item \textbf{\HyperFair (NP + Val):} The \textbf{\PSL \, Base} model with both the non-parity and value fairness rules.
    
    \item \textbf{Fair \PSL:} The \textbf{\PSL \, Base} model with rules introduced by Farnadi et al. \cite{fairHyper:RecSys}.
\end{itemize}

We also implemented three of the fair matrix factorization methods introduced by Yao and Huang \cite{DBLP:journals/corr/YaoH17} using the hyperparameters and training methods chosen by the authors, i.e., we use a L2 regularization term $\lambda = 10^{-3}$ and learning rate of $0.1$ for $500$ iterations of Adam optimization using the full gradient.
The first is the baseline matrix factorization based approach which we refer to as \textbf{MF}. 
The second and third methods are where the matrix factorization objective function is augmented with the smoothed non-parity and value unfairness metrics, which we refer to as \textbf{MF NP} and \textbf{MF Val}, respectively.
For both the \HyperFair \, models we introduced in this work and the matrix factorization methods, we set the fairness regularization parameter to $1.0$.


We measure the prediction performance of each of the models using the RMSE of the rating predictions.
The unfairness of the predictions are measured using the metrics defined in \secref{sec:models}.
The prediction performance and fairness metrics are measured across 5 folds of the \MovieLens \, dataset and we report the mean and standard deviation for each model.
We bold the best value, and values not statistically different from the best with $p < 0.05$ for a paired sample t-test.

\begin{table*}[t]
\caption{Prediction performance(RMSE) and unfairness(Non-Parity and Value) of recommender systems on \MovieLens 1m} 
\centering
\begin{tabular}{|l|c|c|c|c|}
\hline
\textbf{Model} & \textbf{RMSE (SD)} & \textbf{Non-Parity (SD)} & \textbf{Value (SD)}\\ 
 \hline
 \hline
 MF & $0.945 (1.0e\hbox{-}3)$ & $0.0371(1.6e\hbox{-}3)$ & $0.349 (6.0e\hbox{-}3)$ \\
 \hline
 MF NP & $0.945(1.0e\hbox{-}3)$ & $\mathbf{0.0106(2.0e\hbox{-}3)}$ & $0.351(6.3e\hbox{-}3)$ \\
 \hline
 MF Val & $0.950(6.4e\hbox{-}4)$ & $0.0446(3.0e\hbox{-}3)$ & $0.343(3.4e\hbox{-}3)$ \\
 
\hline
\hline
Fair \PSL & $\mathbf{0.932(9.7e\hbox{-}4)}$ & $0.0274(9.6e\hbox{-}4)$ &  $\mathbf{0.332(5.5e\hbox{-}3)}$ \\ 
 
 \hline
 \hline
PSL Base & $\mathbf{0.931 (1.2e\hbox{-}3)}$ & $0.0270 (1.4e\hbox{-}3)$ & $\mathbf{0.330 (4.4e\hbox{-}3)}$ \\ 
\hline 
\HyperFair (NP) & $0.945 (1.1e\hbox{-}2)$ & $0.0215 (5.0e\hbox{-}3)$ & $\mathbf{0.338(9.9e\hbox{-}3)}$ \\ 
\hline 
\HyperFair (Val) & $\mathbf{0.932(1.1e\hbox{-}3)}$ & $0.0267(8.6e\hbox{-}4)$ & $\mathbf{0.333(6.9e\hbox{-}3)}$ \\
\hline
\HyperFair (NP + Val) & $\mathbf{0.932 (1.1e\hbox{-}3)}$ & $0.0274(1.4e\hbox{-}3)$ & $\mathbf{0.331(4.5e\hbox{-}3)}$ \\
\hline
\end{tabular}
\label{tab:fairness_results}
\end{table*}

We can see from \tabref{tab:fairness_results} that the \HyperFair \, models either improved on the targeted fairness metric over \textbf{\PSL \, Base} or achieved results that were not significantly different from the best \PSL \, model.
Notably, \textbf{\HyperFair(NP)} achieved significantly better non-parity unfairness over \textbf{\PSL \, Base} and the anticipated performance decrease in the RMSE of the rating predictions was minimal.
In fact, the \textbf{\HyperFair(NP)} still achieved the same level of prediction accuracy as the highest performing matrix factorization method. 
Another interesting takeaway from \tabref{tab:fairness_results} is that when attempting to optimize for both non-parity and value unfairness simultaneously in \textbf{\HyperFair(NP + Val)}, the effectiveness of the non-parity rule decreases.
This effect can also be observed in the \textbf{ Fair \PSL} of \cite{farnadi2018fairness}, where both fairness notions are attempting to be addressed with a single set of rules.
A potential explanation of this could be that the value unfairness and non-parity unfairness metrics are opposing one another, that is to say that a set of ratings that performs well on value unfairness in this dataset may actually perform poorly on non-parity unfairness, and vice-versa.
Fully understanding this behavior and controlling the tradeoff between the metrics is a direction for future work.

When comparing the \PSL \, models to the matrix factorization models from \cite{DBLP:journals/corr/YaoH17}, we see that \PSL \, consistently achieves better RMSE and value unfairness, while matrix factorization achieves better non-parity unfairness values.
It is important to note that \tabref{tab:fairness_results} only reflects metrics values for the regularization parameter $w_f = 1.0$.
In the next set of experiments, we show how tuning the non-parity unfairness regularization parameter can effectively yield predictions that fall within a desired non-parity unfairness threshold.

\subsection{Post-Process Fairness Intervention} \label{post_process_exp_section}
Another way we employ our proposed methods is as an interpretable fair retrofitting procedure for predictions from an arbitrary black-box model.
To show the effectiveness of our methods for this task, we create a simple \PSL \, model that contains only the NMF local predictor rule and the fairness rule.
We refer to these models as \textbf{NMF + NP \PSL} and \textbf{NMF + Val \PSL} for the models with non-parity and value unfairness rules, respectively.
The weights of the fairness rules in the templates are varied to show the tradeoff between prediction performance and the fairness metric.
For both models, we run experiments for all $w_f \in \{0.0, 0.01, 0.1, \cdots, 10000.0 \}$.

\begin{figure}[t]
    \centering
    \subfloat[]{\includegraphics[width=0.4\textwidth]{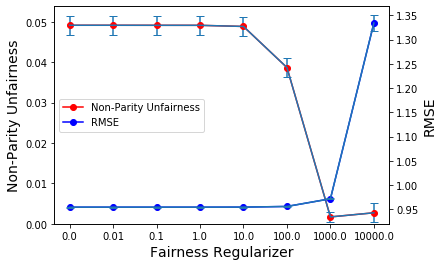}}\hfill
    \subfloat[]{\includegraphics[width=0.4\textwidth]{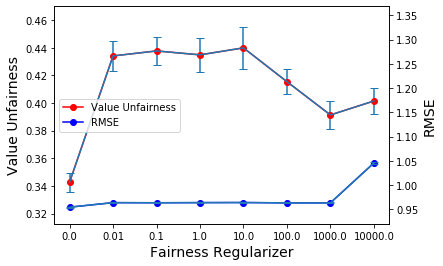}}
    \caption{(a) Non-Parity unfairness and RMSE performance of \textbf{NMF + NP \PSL} vs the value of the fairness regularization parameter.\\ (b) Value unfairness and RMSE performance of \textbf{NMF + Val \PSL} vs the value of the fairness regularization parameter.}
    \label{fig:FairnessvsRegularizer}
\end{figure}
\figref{fig:FairnessvsRegularizer} shows both the RMSE and the fairness of the ratings predicted by the NMF model. 
We see that \textbf{NMF + NP \PSL} begins to improve the predictions' fairness without significantly decreasing the performance when the regularization parameter is set to $100.0$. 
When the parameter exceeds $10.0$, there is a significant decrease in the non-parity unfairness, reaching nearly $0.0$, while the increase in RMSE is still not drastic. 

The \textbf{Val + NP \PSL} model initially does not improve the value unfairness of the NMF rating predictions. 
We suspect this behavior suggests that the quality of the estimator for the group average item rating needs improvement and initially biases the predictions in a detrimental way and is a direction for future investigation.
There is a region where the value unfairness begins a downward trend with respect to the fairness regularizer, as is desired.
This is an encouraging result, showing that the weight of the fair rule generally
has the desired relationship with value unfairness.

%% file: sections/conclusion.tex
We introduced the \HyperFair \, framework for enforcing multiple soft fairness constraints in a hybrid recommender system.
The effectiveness of \HyperFair \, methods are tested in a movie recommendation setting and it is shown they can improve the fairness of predictions and still achieve state-of-the-art performance. 
One direction for future work would be to integrate fairness into the weight learning objective of \PSL.
The experimental results of \secref{post_process_exp_section} showcased that the fairness regularizer is an important hyperparameter to tune, and one way to automate this procedure is via weight learning \cite{bach:jmlr17}.
Another direction for future work is to integrate more fairness metrics and explore what new modeling patterns can be developed for ranking based recommender systems.



%% file: sections/acknowledgements.tex
This work was partially supported by the National Science Foundation grants CCF-1740850 and IIS-1703331, AFRL and the Defense Advanced Research Projects Agency. We also thank the reviewers for their constructive feedback that helped shape the paper.

%% file: sections/appendix.tex

\section{Appendix}
\label{sec:appendix}

\subsection{Probabilistic Soft Logic}

Probabilistic soft logic (\PSL) is a statistical relational learning (SRL) framework that uses arithmetic and first order like logical syntax to define a specific type of probabilistic graphical model called a hinge-loss Markov random field (HL-MRF) \cite{bach:jmlr17}.
To do this, \PSL \, derives potential functions from the provided rules which take the form of hinges.
Data is used to instantiate several potential functions in a process called grounding.
The resulting potential functions are then used to define the HL-MRF.
The formal definition of a HL-MRF is as follows:

\begin{definition}{Hinge-loss Markov random field.}
Let $\mathbf{y} = (y_{1}, \cdots, y_{n})$ be a vector of $n$ variables and $\mathbf{x} = (x_{1}, \cdots, x_{n'})$ a vector of $n'$ variables with joint domain $\mathbf{D} = [0, 1]^{n + n'}$. Let $\mathbf{\phi} = (\phi_{1}, \cdots , \phi_{m})$ be a vector of $m$ continuous potentials of the form $\phi_{i}(\mathbf{y}, \mathbf{x}) = (\max\{\ell_{i}(\mathbf{y}, \mathbf{x}), 0\})^{p_{i}}$,
where $\ell_{i}$ is a linear function of $\mathbf{y}$ and $\mathbf{x}$ and $p_{i} \in \{1,2\}$. 
Let $\mathbf{c} = (c_{1}, \cdots, c_{r})$ be a vector of $r$ linear constraint functions associated with index sets denoting equality constraints $\mathcal{E}$ and inequality constraints $\mathcal{I}$, which define the feasible set 
.
\begin{equation*}
\Tilde{\mathbf{D}} 
= \left\{
(\mathbf{y},\mathbf{x}) \in \mathbf{D} \, \Bigg  \vert 
\begin{array}{lr}
        c_{k}(\mathbf{y}, \mathbf{x}) = 0,& \forall k \in \mathcal{E}\\
        c_{k}(\mathbf{y}, \mathbf{x}) \leq 0,& \forall k \in  \mathcal{I}\\
\end{array}
\right\}
\end{equation*}
Then, for $(\mathbf{y}, \mathbf{x}) \in \Tilde{\mathbf{D}}$, given a vector of $m$ nonnegative parameters, i.e., weights, $\mathbf{w} = (w_{1}, \cdots, w_{m})$, a \textbf{hinge-loss Markov random field} $\mathcal{P}$ over random variables $\mathbf{y}$ and conditioned on $\mathbf{x}$ is a probability density defined as:

\begin{equation}
    P(\mathbf{y} \vert \mathbf{x}) = 
    \begin{cases}
    \frac{1}{Z(\mathbf{w}, \mathbf{x})} \exp (-\sum_{j = 1}^{m} w_{j} \phi_{j}(\mathbf{y}, \mathbf{x})) & (\mathbf{y}, \mathbf{x}) \in \Tilde{\mathbf{D}} \\
    0 & o.w.
    \end{cases}
    \label{eq:HL-MRF_Dist}
\end{equation}

where 
$Z(\mathbf{w}, \mathbf{x}) = \int_{\mathbf{y} \vert (\mathbf{y}, \mathbf{x} \in \Tilde{\mathbf{D}})} \exp(-f_{\mathbf{w}}(\mathbf{y}, \mathbf{x})) d\mathbf{y}$
is the partition function for the conditional distribution.

\end{definition}

Rules in a \PSL \, model capture interactions between variables in the domain and can be in the form of a first order logical implication or a linear arithmetic relation.
Each rule is made up of predicates with varying numbers of arguments.
Substitution of the predicate arguments with constants present in the data generate ground atoms that can take on a continuous value in the range $[0, 1]$.
A logical rule must have a conjunctive clause in the body and a disjunctive clause in the head, while an arithmetic rule must be an inequality or equality relating two linear combinations of predicates.

A logical rule is translated as a continuous relaxation of Boolean connectives using \textit{Lukasiewicz} logic.
Specifically, $P \psland Q$ results in the potential $\max(0.0, P \pslsum Q - 1.0)$, $P \pslor Q$ results in the potential $\min(1.0, P \pslsum Q)$, and $\pslneg Q$ results in the potential $1.0 - Q$.
Each grounding of an arithmetic rule is manipulated to $\ell(\mathbf{y}, \mathbf{x}) \leq 0$ and the resulting potential takes the form $\max \{\ell(\mathbf{y}, \mathbf{x}), 0\}$.

We now illustrate the process of instantiating a HL-MRF using \PSL \, with an example in the context of recommender systems.

\begin{center}
\begin{tabular}{ |l|r| } 
 \hline
 $\pslpred{SimUser}(\pslarg{U1},\pslarg{U2}) \psland \pslpred{Rating}(\pslarg{U1}, \pslarg{M})
    \pslthen \pslpred{Rating}(\pslarg{U2}, \pslarg{M})$ & (1) \\ 
 $\pslneg \pslpred{SimUser}(\pslarg{U1}, \pslarg{U2}) \pslor \pslneg \pslpred{Rating}(\pslarg{U1}, \pslarg{M}) 
    \pslor \pslpred{Rating}(\pslarg{U2}, \pslarg{M})$ & (2) \\ 
 $min\{1.0, (1.0 - \pslpred{SimUser}(\pslarg{Alice}, \pslarg{Bob}) \pslsum
    (1.0 - \pslpred{Rating}(\pslarg{Alice}, \pslarg{Alien})) \pslsum
    \pslpred{Rating}(\pslarg{Bob}, \pslarg{Alien})\}$ & (3) \\ 
 \hline
\end{tabular}
\end{center}

\begin{exmp}
Consider a movie recommendation setting where the task is to predict the ratings users will provide movies that they have not yet rated.
We can encode the idea that similar users are likely to enjoy the same movies using the logical statement (1).  

Here, \pslpred{SimUser} is an observed predicate that represents the similarity between two users, and \pslpred{Rating} is the predicate that we are trying to predict, i.e., the rating a user will give a movie.
\PSL \, first converts the rule to its disjunctive normal form, (2). 
Then, all possible substitutions of constants for the variable arguments in the predicates of the rule are made to make ground atoms.
Then, all possible combinations of atoms that can form a ground rule are made.

Finally, by utilizing the \textit{Lukasiewicz} relaxation, the following hinge-loss function is created.
For instance, let us assume we have data with users $\pslarg{U} = \{Alice, Bob\}$ and movies $\pslarg{M} = \{Alien\}$.
This will result in the hinge-loss function (3).
\end{exmp}

We refer the reader to \cite{bach:jmlr17} for a more detailed description of \PSL.